# Lateral Transport and Field-Effect Characteristics of Sputtered P-Type Chalcogenide Thin Films

Sumaiya Wahid[1], Alwin Daus[1], Asir Intisar Khan[1], Victoria Chen[1], Kathryn M. Neilson[1], Mahnaz Islam[1] and Eric Pop[1,2,a)]

[1]Department of Electrical Engineering, Stanford University, Stanford, California 94305, USA
[2]Department of Materials Science and Engineering, Stanford University, Stanford, California 94305, USA

**ABSTRACT:** Investigating lateral electrical transport in p-type thin film chalcogenides is important to evaluate their potential for field-effect transistors (FETs) and phase-change memory applications. For instance, p-type FETs with sputtered materials at low temperature ($\leq 250°C$) could play a role in flexible electronics or back-end-of-line (BEOL) silicon-compatible processes. Here, we explore lateral transport in chalcogenide films ($Sb_2Te_3$, $Ge_2Sb_2Te_5$, $Ge_4Sb_6Te_7$) and multilayers, with Hall measurements (in $\leq 50$ nm thin films) and with p-type transistors (in $\leq 5$ nm ultrathin films). The highest Hall mobilities are measured for $Sb_2Te_3$/GeTe superlattices (~18 $cm^2V^{-1}s^{-1}$ at room temperature), over 2-3× higher than the other films. In ultrathin p-type FETs with $Ge_2Sb_2Te_5$, we achieve field-effect mobility up to ~5.5 $cm^2V^{-1}s^{-1}$ with current on/off ratio ~$10^4$, the highest for $Ge_2Sb_2Te_5$ transistors to date. We also explore process optimizations (e.g., $AlO_x$ capping layer, type of developer for lithography) and uncover their trade-offs towards the realization of p-type transistors with acceptable mobility and on/off current ratio. Our study provides essential insights into the optimization of electronic devices based on p-type chalcogenides.

Sputtered tellurium (Te) based chalcogenides with germanium (Ge) and antimony (Sb) have attracted attention for various electronic applications such as phase-change memory, p-type transistors, and thermoelectric applications.[1-5] A key advantage of these materials is their low deposition and low crystallization temperature which makes them compatible with flexible electronics ($\leq 250°C$ deposition temperature[6]) and heterogeneous integration with silicon back-end-of-line (BEOL) applications (thermal budgets below 2 hours at ~500°C[7]). That these materials have p-type semiconducting properties makes them particularly interesting because of difficulties in finding inorganic p-type material candidates deposited with low-cost, large-area compatible methods.[8,9]

Lateral (in-plane) transport studies of bulk crystalline films of these chalcogenides using Hall effect measurement demonstrate good hole mobility at room temperature: ~350-400 $cm^2V^{-1}s^{-1}$ for $Sb_2Te_3$,[10,11] ~120 $cm^2V^{-1}s^{-1}$ for GeTe,[12,13] and ~30-50 $cm^2V^{-1}s^{-1}$ for $Ge_2Sb_2Te_5$.[12,14] There are few reports involving $Ge_2Sb_2Te_5$ thin film (10-100 nm) transistors that show limited performance, not reaching the high mobility found in bulk and an on/off current ratio less than 20.[2,15-19] Recently, 5 nm thick $Ge_2Sb_2Te_5$ was used in flexible thin film transistors achieving an improved on/off ratio up to ~388.[20,21] On the other hand, recent developments in phase-change memory technology have made use of chalcogenide superlattice structures consisting of layers each only a few nanometers thick.[22]

In this work, we investigate the lateral electrical transport properties of ~50 nm thick single layer Ge, Sb, and Te-based chalcogenide compounds, which include $Sb_2Te_3$, $Ge_2Sb_2Te_5$, and $Ge_4Sb_6Te_7$, a relatively new composition.[23] We compare their properties with multilayer films (23 nm and 53 nm total thickness) consisting of $Sb_2Te_3$/GeTe (4 nm/1 nm) superlattices[24] using Hall effect and transfer length method (TLM)

---

measurements. We also fabricate transistors with ultrathin (4 to 5 nm) films of $Sb_2Te_3$, $Ge_2Sb_2Te_5$ and $Ge_4Sb_6Te_7$, as well as bilayer $Sb_2Te_3/Ge_2Sb_2Te_5$ (2 nm/3 nm); all sputtered at ≤ 180ºC, drawing a comparison among different materials as well as devices with or without *ex situ* $AlO_x$ capping. In doing so, we determine the optimized device performance in terms of field-effect mobility (and thus on-state current) and the gate modulation capability (current on/off ratio) of the films.

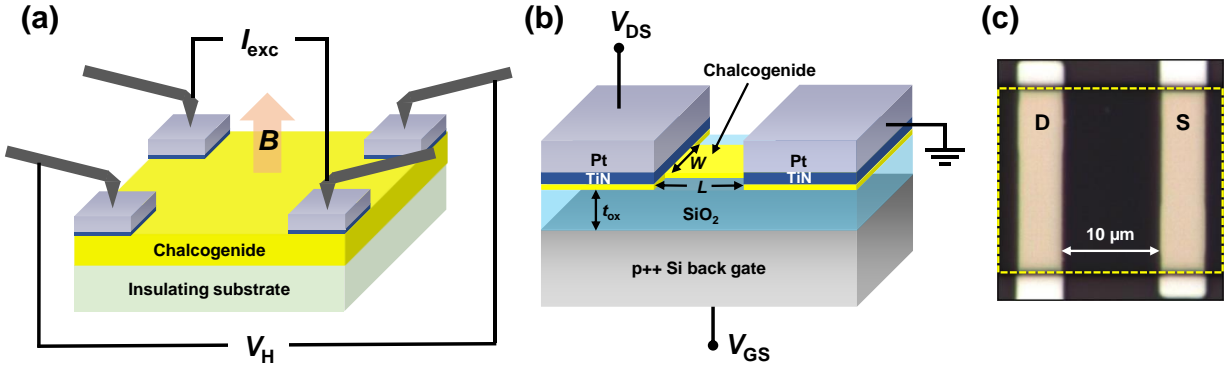

FIG. 1. (a) Schematic of the fabricated van der Pauw Hall measurement structure. (b) Schematic of back-gated transistor structure for field-effect measurements. (c) Top view optical image of a fabricated transistor. Yellow dashes mark the outline of the chalcogenide channel, here with $L$ = 10 μm and $W$ = 20 μm. $I_{exc}$ = excitation current, $V_H$ = Hall voltage, $B$ = magnetic field, $V_{DS}$ = drain-source voltage, $V_{GS}$ = gate-source voltage, $L$ = channel length, $W$ = channel width, $t_{ox}$ = oxide thickness, D = drain, S = source.

We start by performing Hall measurements on ~50 nm thin films of $Sb_2Te_3$, $Ge_2Sb_2Te_5$, and $Ge_4Sb_6Te_7$, as well as multilayer $Sb_2Te_3$/GeTe stacks, before fabricating back-gated transistors with ultrathin films of these materials. A schematic of the van der Pauw Hall measurement structure is shown in Fig. 1(a), while the schematic and top view optical image of the fabricated transistors are shown in Figs. 1(b) and 1(c), respectively. The results from Hall effect measurement are summarized in Table I. We also determine the mobility and carrier concentration of the multilayer samples using TLM and find similar values compared to the Hall measurements. The Hall and TLM sample fabrication and measurement parameters are described in the Supplementary Material (Sections S1 and S2). We plot these results of mobility vs. carrier concentration together with reported literature[20,25-27] on similar superlattices ($Sb_2Te_3$/GeTe or $Sb_2Te_3/Ge_2Sb_2Te_5$) as well as $Sb_2Te_3$, $Ge_2Sb_2Te_5$, and $Ge_6Sb_1Te_2$ thin films (≤ 100 nm) in Fig. 2. This comparison reveals significant improvement in mobility of multilayer films compared to the single layer chalcogenides. This can be attributed to the ordering of vacancies and formation of van der Waals-like gaps in the multilayer films,[24,25,28] which can enhance the lateral transport due to improved interfaces.

TABLE I. Hall measurement data for p-type polycrystalline chalcogenide thin films and superlattices. All measurements are at room temperature in air. Annealing was also performed in air.

| Material | Thickness (nm) | Annealing (A) or Deposition (D) Temperature (ºC) | Carrier Type | Carrier Concentration ($cm^{-3}$) | Hall Mobility ($cm^2V^{-1}s^{-1}$) | Resistivity (Ω·cm) |
|---|---|---|---|---|---|---|
| $Sb_2Te_3$ | 50 | 180 (A) | p | $5.1 \times 10^{20}$ | 8.5 | $1.5 \times 10^{-3}$ |
| $Ge_2Sb_2Te_5$ | 50 | 180 (A) | p | $2.6 \times 10^{20}$ | 6.4 | $3.8 \times 10^{-3}$ |
| $Ge_4Sb_6Te_7$ | 50 | 200 (A) | p | $8.4 \times 10^{20}$ | 1.2 | $6.4 \times 10^{-3}$ |
| $Sb_2Te_3$/GeTe | 53 | 180 (D) | p | $2.2 \times 10^{20}$ | 16.3 | $1.8 \times 10^{-3}$ |
| $Sb_2Te_3$/GeTe | 23 | 180 (D) | p | $2.1 \times 10^{20}$ | 18.0 | $1.7 \times 10^{-3}$ |

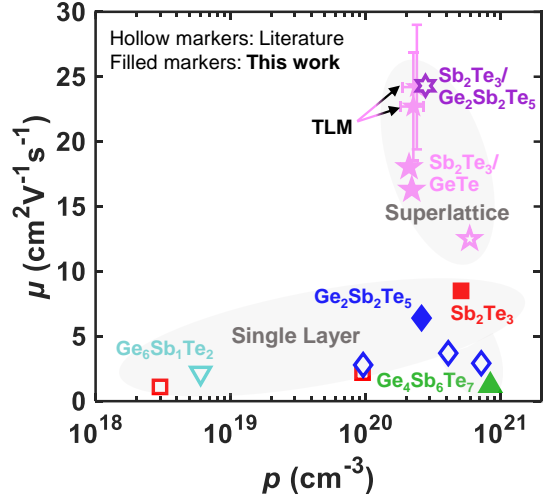

FIG. 2. Plot of hole mobility ($\mu$) vs. concentration ($p$) for single layer and superlattice chalcogenides with $\leq$ 100 nm film thickness.[20,25-27] All data are from Hall measurements, unless otherwise indicated. Error bars in the TLM data represent standard deviation among 4 TLMs with at least 6 different channel lengths.

The Hall measurements also determine p-type conductivity with very high carrier concentration in all thin films, possibly due to high concentration of Ge or Sb vacancies (in $Ge_2Sb_2Te_5$)[26,29] or $Sb_{Te}$ anti-site defects (in $Sb_2Te_3$)[30] for the crystalline phases of these materials, thus forming p-type degenerate semiconductors. The results are in line with reported literature on $Ge_2Sb_2Te_5$ and $Sb_2Te_3$ films [as shown in Fig. 2]. In this work, we also perform the first Hall measurements on the newly-discovered[23] composition $Ge_4Sb_6Te_7$. This material exhibits higher hole concentration, which could be attributed to the lower band gap of crystalline $Ge_4Sb_6Te_7$, and lower Hall mobility which could be due to the co-existence of mixed phase (Ge-Sb-Te and Sb-Te) regions.[23] However, we note that the resistivity (i.e. the inverse of the product of mobility and hole concentration) of the $Ge_4Sb_6Te_7$ film is only ~1.7× higher than that of $Ge_2Sb_2Te_5$ deposited and annealed under similar conditions, as shown in Table I.

Unfortunately, for field-effect transistor applications [Figs. 1(b) and 1(c)], the transistor channel with such high carrier concentrations cannot be easily depleted, which prevents appreciable gate modulation in films with tens of nanometers thickness. In this case, fully depleting the semiconductor with the gate-induced electric field to "turn off" such transistors, requires an ultrathin semiconductor channel. Thinning down the material has proven effective for improved gate modulation[20] partially aided by a possible increase in its band gap at sub-5 nm thickness.[31] Thus, we investigate the behavior of chalcogenide transistors at a reduced thickness of 4 to 5 nm. We fabricate both single and bilayer channel transistors of two types: uncapped and *ex situ* Al-capped (naturally oxidized to $AlO_x$), where the latter prevents damage to the chalcogenide during contact patterning in optical ultraviolet (UV) lithography (for details see Supplementary Material Sections S1 and S3). The devices are annealed at or above their respective crystallization temperatures (~110°C for $Sb_2Te_3$,[4] ~150°C for $Ge_2Sb_2Te_5$,[1] and ~200°C for $Ge_4Sb_6Te_7$[23]) to ensure polycrystallinity irrespective of their deposition temperatures.

Figure 3 displays measured drain current ($I_D$) vs. gate voltage ($V_{GS}$) for the devices with 10 µm channel length at a drain-source voltage of -1 V. In all cases, the uncapped devices (solid line) have lower on and off-state current ($I_{on} = I_D$ at $V_{GS} = -70$ V; $I_{off} = I_D$ at $V_{GS} = 70$ V) but better gate modulation compared to the AlO$_x$-capped devices (dashed line). The lower $I_{on}$ in the uncapped devices may be attributed to the basic photoresist developer partially etching the semiconductor underneath the contact region, leading to partially degraded contacts in these devices. The devices capped with AlO$_x$, which is intended to circumvent this issue, show improved on-state current, but their current on/off ratio ($I_{on}/I_{off}$) is lowered drastically mainly due to higher off-state current compared to their uncapped counterparts. Possible reasons for this will be discussed below.

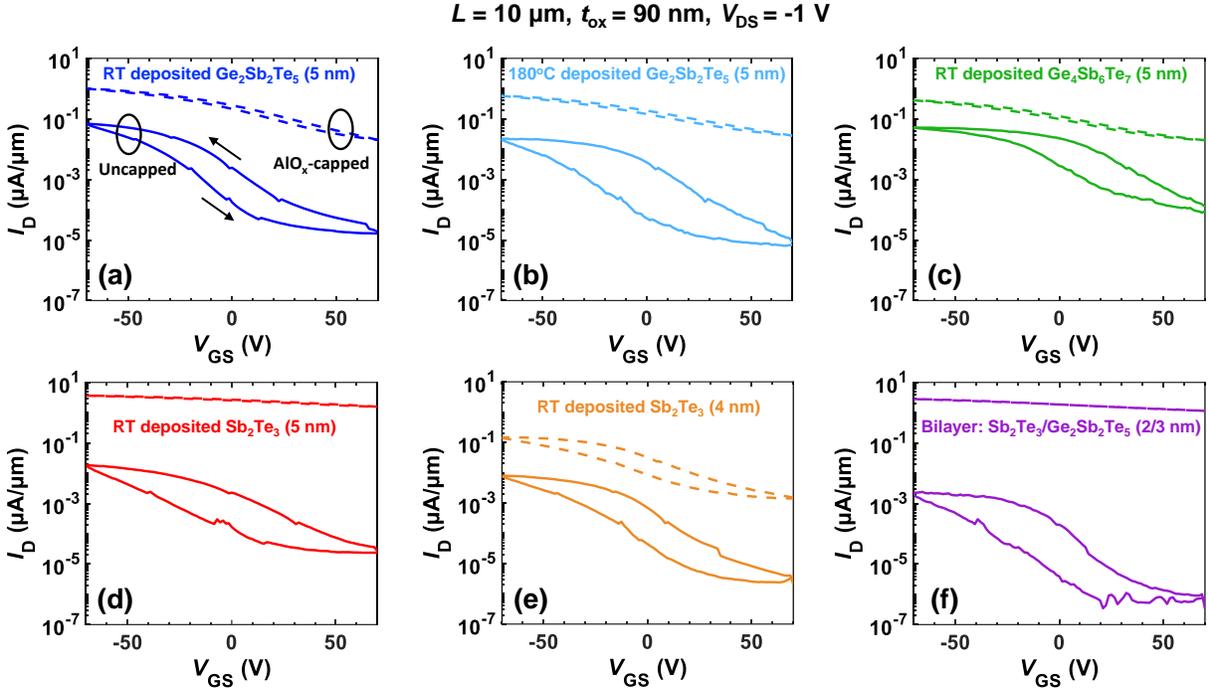

FIG. 3. Measured transfer characteristics ($I_D$ vs. $V_{GS}$) of uncapped (solid line) and AlO$_x$-capped (dashed line) 5 nm thick (a) Ge$_2$Sb$_2$Te$_5$ deposited at room temperature (RT), (b) Ge$_2$Sb$_2$Te$_5$ deposited at 180ºC, (c) Ge$_4$Sb$_6$Te$_7$ deposited at RT, (d) Sb$_2$Te$_3$ deposited at RT, (d) 4 nm thick Sb$_2$Te$_3$ deposited at RT, and (f) 5 nm thick bilayer (2 nm Sb$_2$Te$_3$/3 nm Ge$_2$Sb$_2$Te$_5$) devices. The gate current of all devices was negligible, below ~0.1 nA/μm. Arrows indicate voltage sweep direction, revealing some hysteresis in these uncapped transistors, measured in air. $L$ = channel length, $t_{ox}$ = oxide thickness, $V_{DS}$ = drain-source voltage.

In addition, using a more benign developer can shed more light into the effect of developer solution on the contact region. Both electron-beam lithography (EBL) and deep ultraviolet (DUV) lithography utilize a benign solvent-based developer,[32,33] which is not detrimental to the underlying chalcogenide. As a result, we pattern two of the samples (Ge$_2$Sb$_2$Te$_5$ deposited at 180ºC and bilayer Sb$_2$Te$_3$/Ge$_2$Sb$_2$Te$_5$) with EBL (details in Supplementary Material Section S1) and compare their performance with the devices patterned with UV lithography. The transfer characteristics ($I_D$ vs. $V_{GS}$) and the output characteristics ($I_D$ vs. $V_{DS}$) of the EBL devices patterned with solvent-based developer are shown in Fig. 4. Both devices show similar current on/off ratio as their respective uncapped devices patterned with UV lithography. However, there is a significant improvement in their on-state currents which is ~15x for the Ge$_2$Sb$_2$Te$_5$ deposited at 180ºC. We attribute the improvement in on-state current to the contact region being unaffected by the developer during the contact patterning process. However, we note that EBL poses processing challenges for some flexible substrates, limited by thermal and radiation tolerances and/or unwanted charging effects.[34] EBL is also not an easily scalable manufacturing technique, therefore the use of optical lithography on such chalcogenides should be carefully considered based on the type of substrate, developer solution etc.

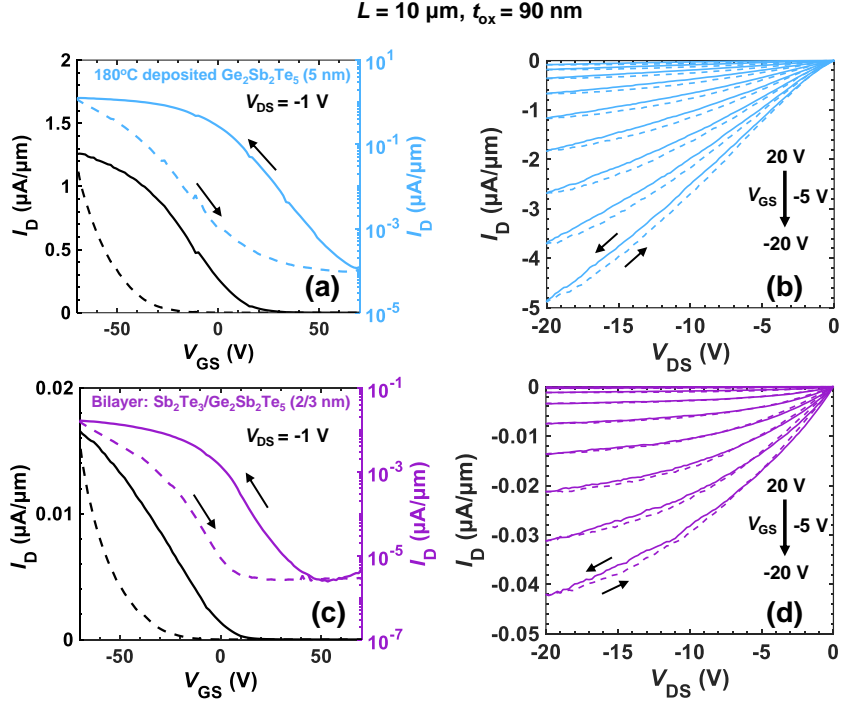

FIG. 4. Measured (a) transfer and (b) output characteristics of uncapped 5 nm thick $Ge_2Sb_2Te_5$ deposited at 180ºC. This represents a 'hero' device, with field-effect mobility $\mu_{FE} \approx 5.5$ cm$^2$V$^{-1}$s$^{-1}$. Measured (c) transfer and (d) output characteristics of a bilayer (2 nm $Sb_2Te_3$/3 nm $Ge_2Sb_2Te_5$) device. Both devices were patterned with electron-beam lithography (EBL). Small arrows represent voltage sweep directions. Note linear and logarithmic scales in (a) and (c).

To gain more insights, Fig. 5 compares the on-state current and field-effect mobility ($\mu_{FE}$) of different materials vs. the current on/off ratio. Overall, the uncapped samples show higher on/off ratio whereas the $AlO_x$-capped samples have lower on/off ratio, as shown by the two shaded regions in Fig. 5. While EBL patterned uncapped samples (triangle symbols) show improved on-state current compared to uncapped devices patterned with UV lithography (circle symbols), the $AlO_x$-capped samples (square symbols) still show significantly higher on-state current. This indicates an additional phenomenon responsible for higher on-state current in the $AlO_x$-capped devices. Here, we observe a simultaneous increase in both on-state current and off-state current indicating higher mobility and/or carrier concentration, which could be attributed to possible Al diffusion into the chalcogenide film[35] or an interfacial doping effect.[36,37]

The effect of reducing the channel thickness is also shown in Fig. 5 with a dashed arrow. We fabricate both uncapped and $AlO_x$-capped $Sb_2Te_3$ samples with thinner sputtered channels (4 nm, 3 nm, and 2 nm) in order to improve the gate modulation. The 4 nm thick $Sb_2Te_3$ sample (shown in Fig. 5) expectedly shows higher on/off ratio but lower on-state current because there is a trade-off when scaling down the thickness. The 3 nm and 2 nm samples showed very low on-state current and hence are not included in the plot. The lowering of mobility is possibly caused by increased carrier scattering at reduced channel thickness.[31] As a result of such degradation at lower sputtered thicknesses, we kept our investigation of other materials down to 5 nm thick channels.

Interestingly, we do not observe improved performance for the bilayer $Sb_2Te_3$/$Ge_2Sb_2Te_5$ transistor compared to its single layer counterparts, as was indicated by the Hall and TLM measurements of thicker (23 nm and 53 nm) superlattice stacks. This could be because the bilayer sample was fabricated with a thickness limited to 5 nm in order to achieve appreciable gate modulation. It should be noted that the bilayer and superlattice structures are fabricated by first depositing a seed layer ($Sb_2Te_3$) at room temperature before the subsequent chalcogenide deposition(s) at high temperature. This leads to highly oriented crystalline layers with van der Waals-like gaps as reported in literature.[22,24,38] As a result, the multilayer stacks can have improved interfaces between the high-temperature deposited chalcogenide layers, following the first interface with the seed layer. On the other hand, the only interface in the bilayer sample is between the seed layer and the high-temperature deposited layer. The top layers and/or interfaces could be responsible for facilitating the lateral transport in the thicker superlattices and thus the improvement is not observed in the bilayer sample. Moreover, the superlattice stack is formed with thicker $Sb_2Te_3$ layers (both as a seed layer and in subsequent sub-layers) in contrast to using only 2 nm $Sb_2Te_3$ seed layer in the bilayer sample, which could have a distinct role in crystallization of the subsequent layer(s).[38]

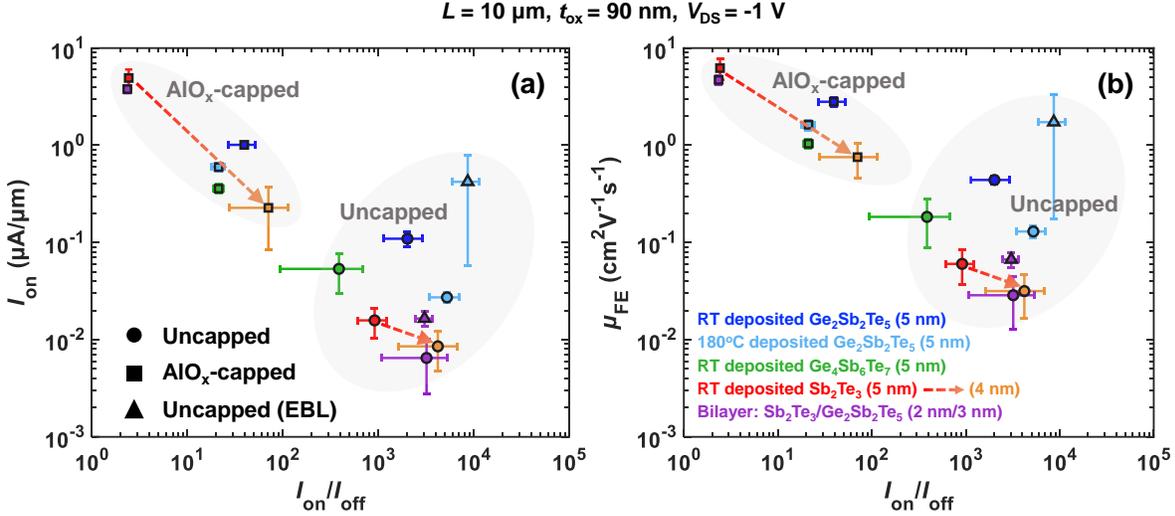

FIG. 5. Plot of our measured (a) on-state current ($I_{on}$) and (b) field-effect mobility ($\mu_{FE}$) vs. current on/off ratio ($I_{on}/I_{off}$) for various fabricated transistors. RT = room temperature, EBL = electron-beam lithography; Error bars represent standard deviation among 5 to 10 devices.

Overall, we achieve the most optimized performance for our p-channel transistors with 5 nm $Ge_2Sb_2Te_5$ deposited at 180°C, patterned with benign solvent-based developer. Our 'hero' device reaches $I_{on} \approx 1.26$ µA/µm at $V_{DS} = -1$ V and $\mu_{FE} \approx 5.5$ cm$^2$V$^{-1}$s$^{-1}$, with $I_{on}/I_{off} \approx 10^4$, and is shown in Figs. 4(a) and 4(b). The average devices have $I_{on} \approx 0.42$ µA/µm at $V_{DS} = -1$ V and $\mu_{FE} \approx 1.7$ cm$^2$V$^{-1}$s$^{-1}$, with $I_{on}/I_{off} \approx 8640$, and the data measured on all such devices are summarized in Supplementary Fig. S3. (All transistors have $L = 10$ µm and are not limited by their contact resistance.[39]) The only other study in the literature which reports Ge, Sb, Te-based chalcogenide thin film transistors below 10 nm thickness is for as-deposited (amorphous) $Ge_2Sb_2Te_5$ films, which show on-state current and field-effect mobility of ~0.01 µA/µm and ~0.04 cm$^2$V$^{-1}$s$^{-1}$, respectively, with $I_{on}/I_{off}$ up to 388.[20,21] Notably, their annealed (crystalline) films at 5 nm thickness showed deteriorated performance, which was attributed to a degradation of the contact and surface area during *ex situ* annealing.[20] In comparison, our hero (average) device shows ~140× (~40×) improvement of field-effect mobility, and nearly 30× improvement in current on/off ratio.

In conclusion, we have shown it is possible to obtain p-channel transistors with mobility 5.5 cm$^2$V$^{-1}$s$^{-1}$ and current on/off ratio ~$10^4$ at room temperature, using 5 nm thick $Ge_2Sb_2Te_5$ films sputtered at 180°C, a deposition process which is compatible with both flexible substrates and BEOL integration. This is important, because there are very few p-channel transistor options,[40-42] as the majority of chalcogenide or oxide transistors are known to be n-type.[43,44] We also find hole Hall mobility up to 18 $cm^2V^{-1}s^{-1}$ in 23 nm thick $Sb_2Te_3$/GeTe superlattices and field-effect mobility up to 6.2 $cm^2V^{-1}s^{-1}$ in 5 nm thick $Sb_2Te_3$ capped with $AlO_x$, at room temperature. These results are also important for further optimization of emerging phase-change memory structures, where both electrical and thermal material properties play a crucial role. Future work must continue to optimize such chalcogenide thin films, e.g. by atomic layer deposition (ALD) or controlled thinning of the channel region of thicker deposited films, which could lead to further improvement in gate modulation with minimal degradation in mobility. The addition of thin high-κ dielectrics or double-gates could also improve the on/off ratio of such p-type transistors.

See supplementary material for details about sample fabrication, measurement, and characterization.

This work was performed at the Stanford Nanofabrication Facility (SNF) and Stanford Nano Shared Facilities (SNSF), supported by the National Science Foundation (NSF) as part of the NNCI under Award No. 1542152. S.W., A.I.K., K.M.N., and M.I. acknowledge the Stanford Graduate Fellowship (SGF) program. A.D. was supported by the Swiss National Science Foundation's Early Postdoc.Mobility fellowship (grant P2EZP2_181619), the Beijing Institute of Collaborative Innovation (BICI) and by the NSF Engineering Research Center for Power Optimization of Electro-Thermal Systems (POETS) with Cooperative Agreement No. EEC-1449548. S.W., A.I.K., and E.P. acknowledge funding from the Semiconductor Research Corporation (SRC) tasks 3004 and 2826, and from the Stanford Non-volatile Memory Technology Research Initiative (NMTRI). V.C. was partly supported by the Stanford SystemX Alliance. E.P. acknowledges partial support from ASCENT, one of six centers in JUMP, an SRC program sponsored by DARPA.

**DATA AVAILABILITY**

The data that support the findings of this study are available from the corresponding author upon reasonable request.

# Supplementary Material

# Lateral Transport and Field-Effect Characteristics of Sputtered P-Type Chalcogenide Thin Films

Sumaiya Wahid[1], Alwin Daus[1], Asir Intisar Khan[1], Victoria Chen[1], Kathryn M. Neilson[1], Mahnaz Islam[1] and Eric Pop[1,2,a]

[1]Department of Electrical Engineering, Stanford University, Stanford, California 94305, USA
[2]Department of Materials Science and Engineering, Stanford University, Stanford, California 94305, USA

## S1. Fabrication Details

I. Van der Pauw Hall Measurement Structures

We fabricate van der Pauw structures for Hall measurements by depositing chalcogenide thin films on insulating substrates (sapphire, or 450 nm $SiO_2$ on Si). For this, we deposit ~50 nm thick single layer films of $Sb_2Te_3$, $Ge_2Sb_2Te_5$ and $Ge_4Sb_6Te_7$ at room temperature by DC and/or RF magnetron sputtering using an AJA ATC 1800-F sputtering system (details in Table S1). Then, we sputter four metal contacts of TiN (10-20 nm)/Pt (40-50 nm) on the corners of the sample using a shadow mask. The films are then annealed *ex situ* for ~30 minutes in air on a hot plate at a temperature of 180ºC for $Sb_2Te_3$ and $Ge_2Sb_2Te_5$ and 200ºC for $Ge_4Sb_6Te_7$, at or above their respective crystallization temperatures.[1-3] Other than films with a single type of chalcogenide, we also investigate the transport properties of multilayer chalcogenide films of $Sb_2Te_3$/GeTe. For this, we first deposit a 3 nm thick $Sb_2Te_3$ seed layer at room temperature, which is annealed *in situ* at 180ºC before depositing alternating stacks of GeTe (1 nm) / $Sb_2Te_3$ (4 nm) at 180ºC without breaking vacuum (~$10^{-7}$ Torr). It has been reported in literature that such seed layers enable the deposition of crystalline and layered films in superlattice stacks.[4-6] We fabricate two samples of such chalcogenide superlattices with total 4 and 10 periods, thus making 23 and 53 nm thick multilayer films, respectively.

TABLE S1. Parameters used for sputtering

| Material | Power (W) | Deposition Pressure (mTorr) | Ar flow (sccm) | Rate (nm/min) |
|---|---|---|---|---|
| $Ge_2Sb_2Te_5$ | 12 (DC) | 2 | 20 | 5.6 |
| $Ge_4Sb_6Te_7$ | 12 (DC) | 2 | 20 | 6.0 |
| $Sb_2Te_3$ | 30 (RF) | 4 | 30 | 1.2 |
| GeTe | 30 (RF) | 4 | 30 | 2.4 |

*sccm: standard cubic centimeters per minute

II. Back-gated Field-Effect Transistor Structures

A schematic diagram illustrating the basic steps of transistor fabrication process is shown in Fig. S1. Single and bilayer material transistors are fabricated on 90 nm thick $SiO_2$ on a degenerately doped p++ silicon substrate which is used as the back-gate electrode. The $SiO_2$ back-gate dielectric is formed through dry thermal oxidation of the substrate. For both uncapped and $AlO_x$-capped devices, we deposit the single layer chalcogenide film (4 to 5 nm) at room temperature or an elevated temperature (180ºC) using DC and/or RF magnetron sputtering (details in Table S1). For the bilayer films, 2 nm $Sb_2Te_3$ is deposited at room temperature as a seed layer, followed by a deposition of 3 nm $Ge_2Sb_2Te_5$ at 180ºC. In case of $AlO_x$- capped devices, an *ex situ* deposition of ~2 nm Al is done on top of the chalcogenide using electron beam evaporation. The deposited Al layer is then allowed to immediately oxidize in air forming an $AlO_x$ capping layer on the semiconductor.

---
[a] Author to whom correspondence should be addressed: epop@stanford.edu

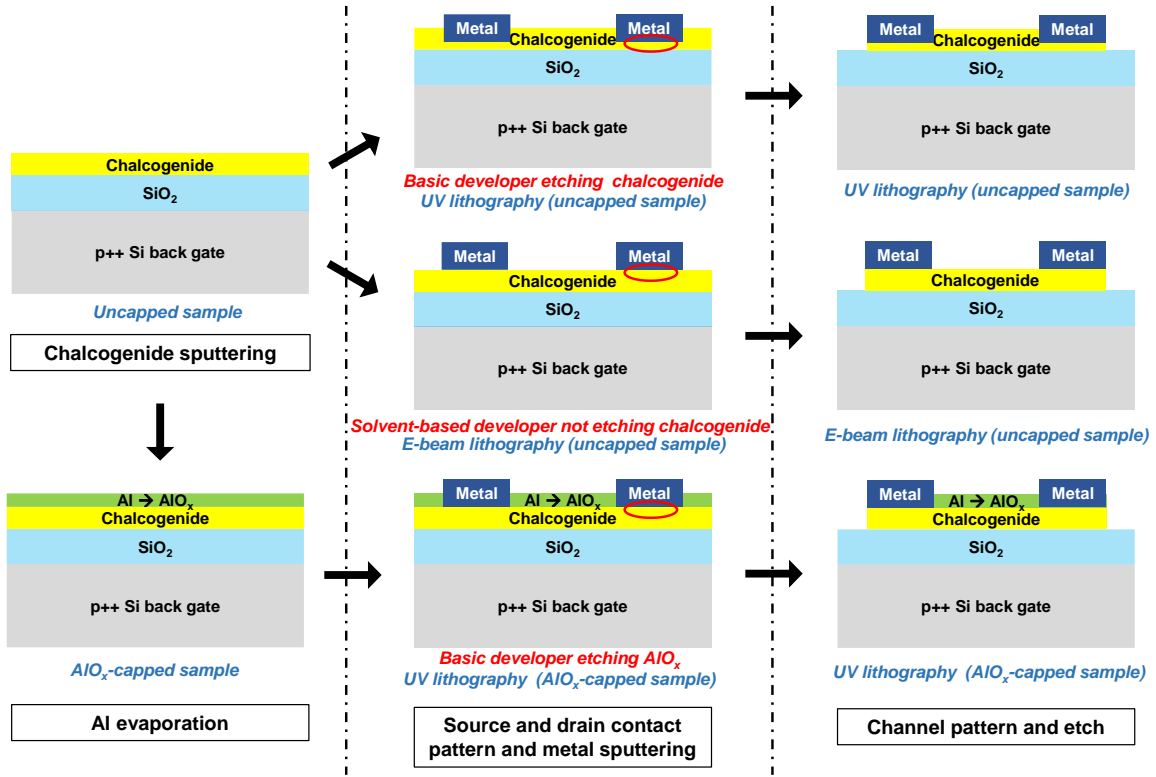

FIG. S1. A schematic illustration of the fabrication process flow.

Next, we define the source and drain metal contacts via optical ultraviolet (UV) lithography in all the samples. In this step, LOL-2000 is used as the lift-off layer (generating an undercut in the resist stack), which is baked at 180ºC for 5 minutes; then, SPR-3612 is used as the photoresist layer (baked at 90ºC for 1 minute). The baking step above the crystallization temperatures of $Sb_2Te_3$ and $Ge_2Sb_2Te_5$ ensures that our investigated films are polycrystalline even in case of a room-temperature deposition. The $Ge_4Sb_6Te_7$ sample, which has a higher crystallization temperature (~200ºC),[3] is later annealed on a hot plate in air at 200ºC for ~30 minutes to ensure polycrystalline film formation. After resist baking, the samples are exposed to 405 nm wavelength laser using a direct write lithography tool (Heidelberg MLA150). Microposit MF-26A, an aqueous developer containing TMAH, is used to develop the pattern in 30-60 seconds, which needs to be handled carefully because such developer can attack the underlying semiconductor film (see Section S3).

For two of the samples (180ºC deposited $Ge_2Sb_2Te_5$ and bilayer $Sb_2Te_3/Ge_2Sb_2Te_5$), we also employ electron-beam lithography (EBL) to define the source and drain contacts. Polymethyl methacrylate (PMMA) 495K A2 (lift-off layer) followed by PMMA 950K A4 form the resist stack (each baked at 180ºC for 5 minutes) for the lift-off process in EBL. The JEOL JBX-6300 electron-beam lithography tool is used to expose the samples to electron beam with a dose of 800 µC cm$^{-2}$, and current of 4 nA (fine contact regions) or 9 nA (etch regions for channel patterning). Then we use a 1:1 solution of methyl isobutyl ketone (MIBK) : isopropyl alcohol (IPA) for 30 seconds for pattern development. This is a solvent-based developer which shows no sign of chalcogenide etching. For all the samples, with the exception of room temperature deposited $Ge_2Sb_2Te_5$, a bilayer metal stack of TiN (10-20 nm) / Pt (40-50 nm) is then sputtered and lifted-off to form the source and drain metal pads. For the room temperature deposited $Ge_2Sb_2Te_5$ an e-beam evaporated Ge/Ni/Au (5 nm/5 nm/30 nm) stack is used as contacts.[7,8] Note, because we investigate long channel devices (10 µm), the transistors are not limited by their contact resistance[9] and the difference in contact materials does not significantly affect our results. Finally, the chalcogenide channel region is patterned with UV lithography and reactive ion etching, 10 mTorr pressure using $CF_4$, $CH_4$ and Ar gases at 60 W bias RF forward power and 400 W ICP RF forward power.

## S2. Electrical Measurements

The Hall measurements are performed using a Lakeshore 8404 Hall Measurement tool, applying an excitation current ($I_{exc}$) of 0.1 mA and a DC magnetic field ($B$) of 0.5 T. The electrical measurements for TLM and transistor devices are done at room temperature in air using a probe station combined with a Keithley 4200-SCS. The TLM measurements were done at $V_{DS}$ = -1 V and the parameters were extracted at $V_{GS}$ = 0 V. In all the TLM and transistor measurements, the gate current did not exceed ~0.1 nA/µm, which is significantly lower than the measured on-state current of the devices.

## S3. Atomic Force Microscopy (AFM)

During the UV lithography process, the thickness of the chalcogenide etched by the MF-26A developer is determined by measuring the step height from the atomic force microscope (AFM) images, captured using Asylum AFM MFP-3D. The numbers, however, may vary because the samples are developed manually, and the rate of agitation or stirring is not identical for all three samples. Nonetheless, because pattern development takes at least 20 to 25 seconds (to wash away the photoresist in the developed regions), the total development time had to be around 30 to 40 seconds to ensure reliable pattern development throughout the chip. During this time, the chalcogenide thickness etched by the developer is ~1.2, 1.6 and 2.1 nm in 30 seconds of total development time for $Ge_2Sb_2Te_5$, $Ge_4Sb_6Te_7$ and $Sb_2Te_3$, respectively.

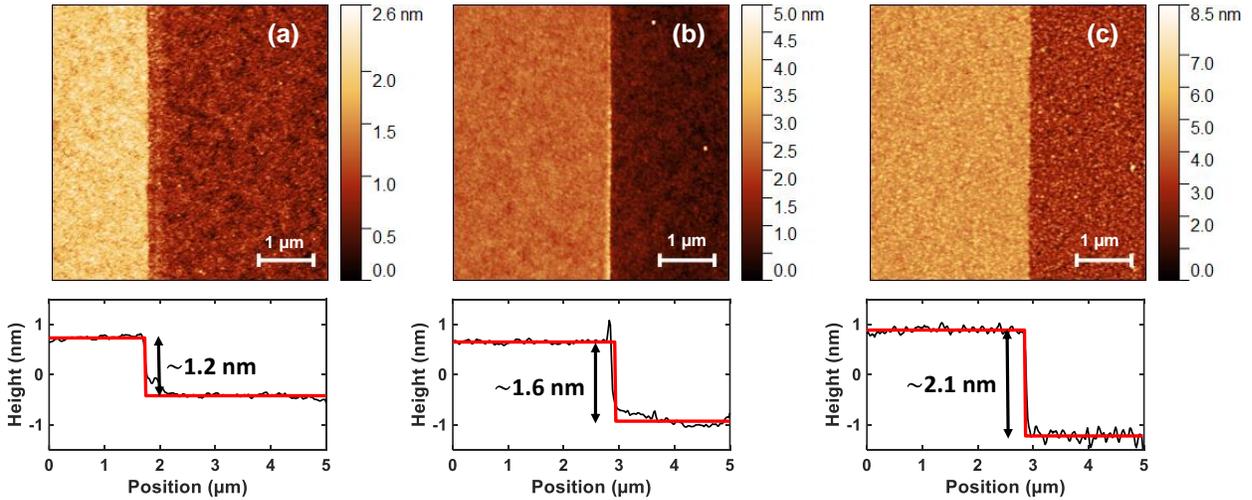

FIG. S2. Atomic force microscope (AFM) images along with respective step heights across the edges of (a) $Ge_2Sb_2Te_5$, (b) $Ge_4Sb_6Te_7$, and (c) $Sb_2Te_3$ samples etched by MF-26A developer.

## S4. Bar graphs of transistor measurements, for Ge$_2$Sb$_2$Te$_5$ deposited at 180°C and patterned with solvent-based developer

The data for field-effect mobility ($\mu_{FE}$), on-state current ($I_{on}$) and current on/off ratio ($I_{on}/I_{off}$) in the 11 measured transistors for Ge$_2$Sb$_2$Te$_5$ deposited at 180°C and patterned with solvent-based developer, are illustrated in Fig. S3. $I_{on}$ refers to $I_D$ at $V_{GS}$ = -70 V and $I_{off}$ refers to $I_D$ at $V_{GS}$ = 70 V. All transistors have $L$ = 10 µm and are measured at $V_{DS}$ = -1 V.

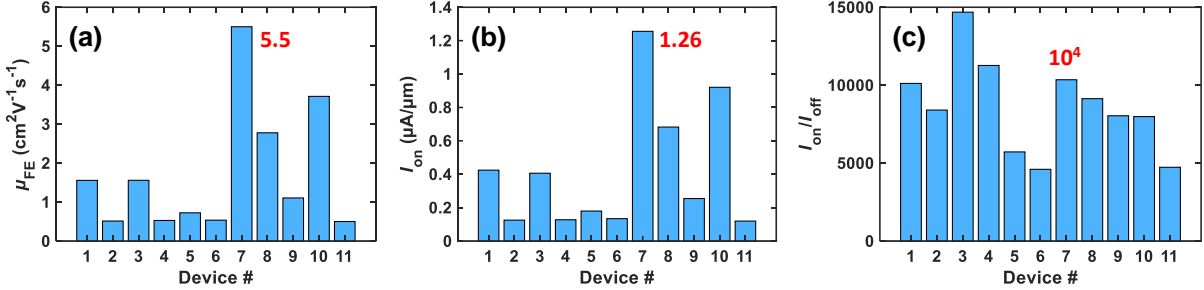

FIG. S3. (a) Field-effect mobility ($\mu_{FE}$), (b) on-state current ($I_{on}$) at $V_{DS}$ = -1 V, and (c) current on/off ratio ($I_{on}/I_{off}$) for different devices in the Ge$_2$Sb$_2$Te$_5$ sample deposited at 180°C and patterned with solvent-based developer. Device # 7 (marked by the red label) represents the 'hero' device mentioned in the main text. The average values are: $\mu_{FE} \approx$ 1.7 cm$^2$V$^{-1}$s$^{-1}$, $I_{on} \approx$ 0.42 µA/µm at $V_{DS}$ = -1 V, and $I_{on}/I_{off} \approx$ 8640.